# Energy Equity, Infrastructure and Demographic Analysis with XAI Methods

Sarahana Shrestha[2], Aparna S. Varde[1,2], Pankaj Lal[2]
1. School of Computing (SoC)
2. Clean Energy and Sustainability Analytics Center (CESAC)
Montclair State University (MSU), NJ, USA

(shresthas1 | vardea | lalp)@montclair.edu
ORCID ID: 0000-0002-9829-9607(Shrestha), 0000-0002-3170-2510 (Varde), 0000-0001-6799-3156 (Lal)

**Abstract**

Understanding the factors influencing energy consumption is crucial to address disparities in energy equity and to promote sustainable solutions. Socio-demographic characteristics can impact energy usage patterns. This study uses XAI methods, e.g. decision trees and PCC, to analyze electricity usage in multiple locales. By correlating the infrastructure and socio-demographic data with energy features, we identify housing tenure and racial demographics as vital predictors, revealing that renters and racially diverse groups can often face higher energy burdens. We demonstrate a novel energy equity web portal and energy burden calculator, hence offering tailored actionable advice to multiple energy stakeholders, along with enhanced explainability. This work addresses challenges in energy policy adaptation, and aims for a next-generation framework, therefore heading more towards energy equity.

**Keywords:** AI in Smart Cities, Energy Burden, Decision Trees, Explainable AI, Pearson's Correlation Coefficient, Sustainable Management, Urban Policy, Web Portal

## Introduction and Related Work

Energy equity aims to ensures that all the communities, and especially marginalized ones, have fair access to affordable energy (Fu et al. 2021). Socio-economic and demographic factors e.g. income, race, housing tenure, housing age, and disparities in infrastructure contribute to inequitable access (Simcock et al. 2021; Singh et al. 2023).

Underserved communities face numerous challenges such as outdated energy infrastructure, inadequate housing, and other structural barriers, thereby limiting their access to energy efficiency programs. Hence, this exacerbates energy burdens and consequently, hinders efforts to transition to more sustainable energy systems (Drehobl et al. 2016). It thus calls for more data-driven analysis with explainable AI (XAI) methods in order to better inform numerous energy policymakers. This motivates our study in this paper.

Related work in the literature (Machlev et al. 2022), (Sim et al. 2022), (Shrestha et al. 2023), (Varde et al. 2023) point out many avenues where XAI methods play vital roles in energy and sustainability analysis. Some works (Conti et al. 2020); (Pawlish et al. 2012) particularly highlight the role of explainable AI paradigms such as commonsense knowledge and decision trees in the context of task efficiency and clean environments respectively, thereby touching upon energy-based aspects. Likewise, other researchers (Basavaraju et al. 2017); (Garg et al. 2020) emphasize the role of XAI based methods in supervised learning for mobile devices, and in generating data for object detection respectively. Both these works address energy tangentially due to their emphasis on making computational processes more efficient. Though we address energy efficiency on the whole for the common good through various policies, it is important to incorporate user opinions (Bittencourt et al. 2024); (Shrestha et al. 2024); (Du et al. 2019), and conduct other relevant analysis to promote more fairness and equity. It is also important to conduct fine-grained analysis in order to derive meaningful inferences, and convey the feedback to stakeholders to make better and more equitable decisions for the future.

We thrive upon many such success stories in the related literature. While these works contribute significantly to the state-of-the-art, we have some originality with respect to our own study. Our work in this paper is novel as being among the first to address energy equity explicitly with a web portal design, energy burden calculator based on demographics and energy infrastructure analysis using XAI methods. This enhances interpretation and trust, hence providing more beneficial feedback to a wide variety of stakeholders.

## Data, Models and Methods

The region of study in this paper is Northeastern United States, more specifically focused on New Jersey (NJ). The Datasets are sourced mainly from NJ programs (New Jersey Clean Energy Program, 2024 Accessed) and the US census from 2008-2022 (U.S. Census Bureau, 2024 Accessed). These are listed next as follows.

- Aggregated Community-Scale Utility Energy Data

- Energy Efficiency Program Participation

- Race (Census Table B02001), Hispanic or Latino Origin (Census Table B03003), Year Structure Built (Census Table B25034), Mean Household Income of Quintiles (Census Table B19081), Household Income (Census Table B19001), Year Structure Built (Census Table B25034), Owner vs. Renter Occupied Units (Census Table B25003)

Explainable AI models are well-deployed in this research. Decision tree classifiers are used to predict the actual energy consumption, quantifying feature importance. Pearson's Correlation Coefficient (PCC) matrix is then used to further assesses these relationships. To calculate energy burden, the total amount spent on energy is divided by the median household income of the locale, as formulated in Equation1.

$$Energy\ Burden\ (\%) = \frac{[(E_e \times R_e) + (E_h \times R_h)]}{M_i} \times 100\% \quad (1)$$

In this equation, $E_e$ is annual household electricity consumption *(kWh)*, $R_e$ is electricity rate *($/kWh)*, $E_h$ is annual heating *(therm/BTU)*, $R_h$ is heating rate *($/therm)*, and $M_i$ is median household income. Note that we use these variables are domain-specific, as typically used.

We design and demonstrate a novel energy web portal with an energy burden calculator. It gets the *user zip-code* as the input, and finds the energy burden using Equation1. If the burden is higher than the state average, it offers clear advice with explainable action items, tailored more towards energy equity. If the burden is below the state average, it mentions that to the users, so they are well aware of their situation. The procedure used in implementing the energy burden calculator is synopsized next in Algorithm1 here.

Using Algorithm1, as well as the analysis with decision trees and Pearson's Correlation Coefficient, several useful insights are obtained in the context of energy equity. In the next section, the results are synopsized with illustration.

*Algorithm 1: Energy Burden Calculator*
**Input**: User Zip-code *($Z_c$)*, State Data *(D)*
**Parameters**: $Z_c$ $E_e$ $R_e$ $E_h$ $R_h$ $M_i$  // domain-specific
**Output**: *EB* (Energy Burden), Message, Display
1: Let *SA = n%*     // state average for EB (constant)
2: Map  $(E_e, E_h)$ → $Z_c$
3: Get  $(R_e, R_h)$  from *D*
4: Compute *E-price* = $E_e * R_e$;   *H-price* = $E_h * R_h$
6: Compute *T-price = E-price + H-price*
7: Compute *EB = (T-price / $M_i$ )*100*
8:      **if** *(EB > n)* **then**
9:       Message = "Overburdened";
10:      Display = Link → {Tips to lower energy burden}
11:      **else** Message = "Below State Average";
12: **return** Print (*EB%*), Print (Message), [Display]

## Results with Discussion

We conduct experiments using decision trees and Pearson's Correlation Coefficient. The results obtained with our XAI classifier models are: *$R^2$=0.7, RMSE=2.5*, implying a good fit of the model to the data for the purpose if prediction. It identifies housing tenure and demographics as being the most significant predictors of electricity consumption. This is observed in Figure 1. In our observations, renter-occupied housing dominates, confirming disparities in energy infrastructure of renters vs. homeowners (Baker et al. 2019). The feature importance of Asian-Americans (15.75%) and owned housing (13.12%) shows homeownership impact on energy equity issues.

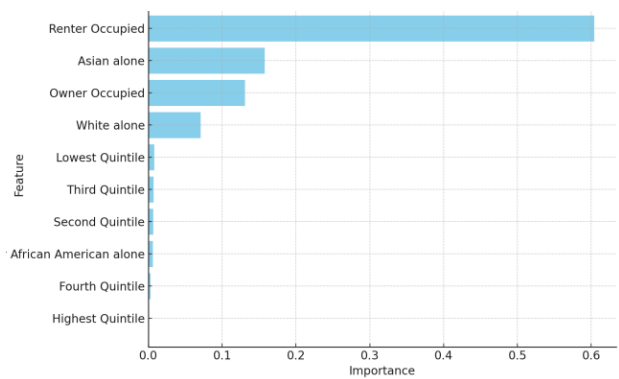

*Figure 1: Feature importance learned from XAI model*

Furthermore, PCC analysis shows that there is a strong correlation between race and home ownership, as observed in Figure 2. Homeownership is more prevalent in the White population (r=0.96), whereas Hispanic or Latino (r=0.93) and Black (r=0.71) populations are more likely to be renters, thus corroborating the systemic disparities noticed in the housing infrastructure (Patterson et al. 2019). Typically, Asian-Americans exhibit moderate positive correlations with new housing (r = 0.52) versus other persons of color (POC), substantiating differences in energy infrastructure (Raymundo 2020). Moreover, the counties with more low-income households tend to have higher rates of renters (r=0.72). Figure 3 shows the PCC matrix between race and income, where we see that the White population shows a strong positive correlation with several income categories, including Low Income (r = 0.93), Moderate Income (r = 0.94), and High Income (r = 0.94) indicating that the White populations are distributed across a wide range of income levels. In contrast, the Hispanic or Latino populations often exhibit a more moderate correlation with Low Income (r = 0.68), which suggests that Hispanic populations are more concentrated in lower-income areas. This suggests that the income disparities exist within this racial group across the different counties (Aliprantis et al., 2024).

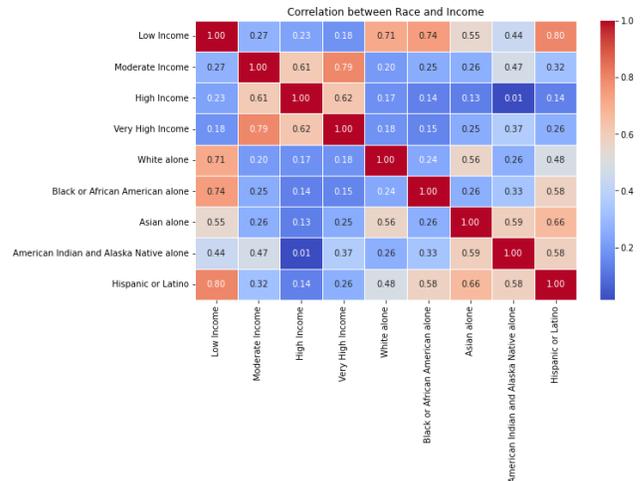

*Figure 3: Pearson Correlation Coefficient (PCC) Matrix for Race and Income*

Based on such results and various other data, Figure 4 presents a few snapshots from the demonstration of our novel web portal for energy management, encompassing an energy burden calculator. These demo snapshots exemplify the useful applications of our XAI analysis on energy equity, targeted for residents, policymakers etc. This is mainly based on infrastructure and demographics.

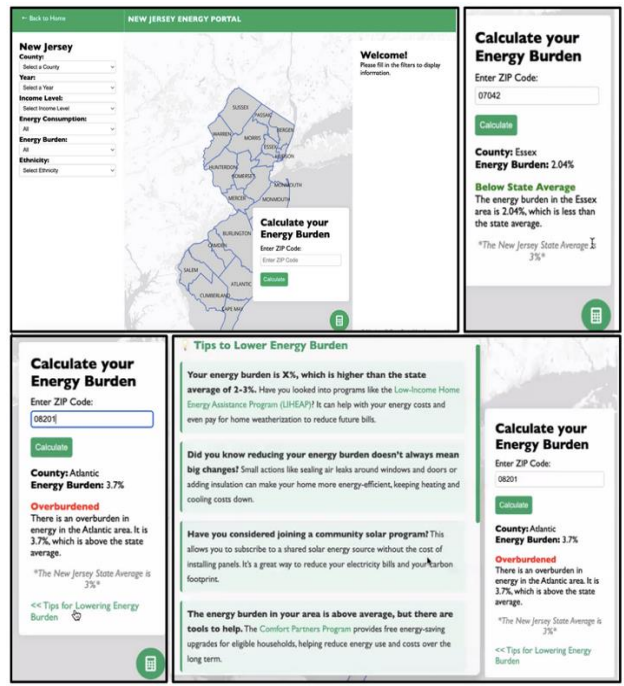

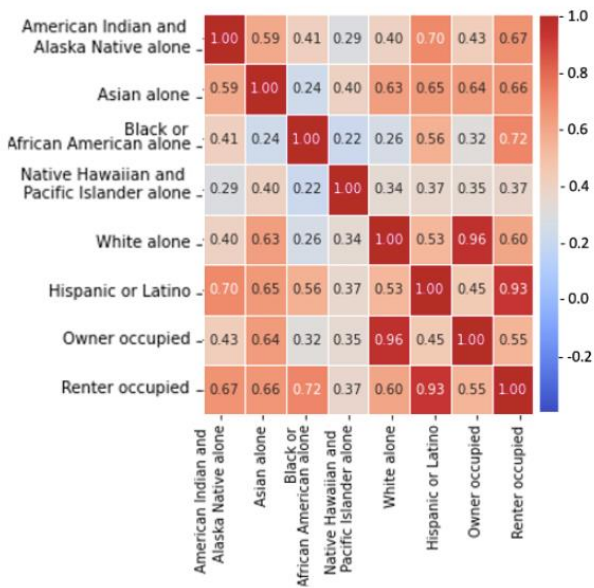

*Figure 2: Pearson Correlation Coefficient (PCC) Matrix for Race and Housing Tenure*

These illustrations in Figures 2 and 3 corroborate many of the observations we noted earlier, thus confirming that our results reveal many significant disparities with respect to energy-related policies and their impacts on the various communities based on the actual demographics as well as the infrastructure. The XAI-based analysis reveals that the alignment of modern infrastructure and affluence are often perpetuating energy inequity.

Figure 4: Energy Management Web Portal and the Energy Burden Calculator Prototype

## Conclusions and Ongoing Work

To the best of our knowledge, ours is among the first works on XAI-based energy equity analysis encompassing a novel energy web portal and energy burden calculator. It addresses crucial challenges in energy adaptation, aiming for higher energy equity, heading towards next-generation systems.

Future work entails: (1) making the energy web portal more interactive to enhance explainability; (2) conducting macro and micro analysis with the countrywide / statewide energy burden; separation of energy sources (gas, solar etc.); aspects such as summer/winter, monthly/annual, and finer granularity in attributes; (3) using methods such as XRRF (explainable reasonably randomized forest) to obtain more robustness & accuracy, yet offering explainable solutions; and (4) enhancing the energy burden calculator further with the expected energy usage analysis in targeted schemes as per the projections. Note that the energy equity analysis based upon decision trees here could be merged with text-based policy perception analysis (Shrestha et al., 2024) in the web portal. Users could compare energy burden data with public opinion trends to see how equity issues align with public attitudes toward renewable energy policies.

It is important to mention that XAI plays a vital role in this research with respect to making all the methods more transparent, and hence interpretable, as well as helping to foster better trust among users. This work on the whole makes broader impacts on AI in Smart Cities and also on the theme of Sustainability Analytics. This is in addition to its main impacts on XAI and Energy & Critical Infrastructure.

## Acknowledgments

Sarahana Shrestha has a Doctoral Assistantship in the PhD Program in Environmental Science and Management from the Graduate School at Montclair State University (MSU). Dr. Aparna Varde acknowledges the NSF MRI grant 2018575, i.e. "MRI: Acquisition of a High-Performance GPU Cluster for Research and Education". The authors thank Gabriel Teves, Julia Konopka, and Joshua Rabo from the Master's Project Teams in the School of Computing at MSU, and Dr. Hao Liu from the Data Science Laboratory, at MSU, for their valuable inputs and useful feedback in this work. We also gratefully acknowledge the travel funding support provided by CSAM, the College of Science and Mathematics at Montclair State University.